\documentclass{sig-alternate-05-2015}
\usepackage{xcolor}
\usepackage{multirow}
\usepackage{caption}
\usepackage{hyperref}
\hypersetup{
  colorlinks,
  citecolor=violet,
  linkcolor=blue,
  urlcolor=cyan}

\begin{document}

\CopyrightYear{2017} 
\setcopyright{acmlicensed}
\conferenceinfo{IWSPA'17,}{March 24 2017, Scottsdale, AZ, USA}
\isbn{978-1-4503-4909-3/17/03}\acmPrice{\$15.00}
\doi{http://dx.doi.org/10.1145/3041008.3041009}







%

\title{Predicting Exploitation of Disclosed Software Vulnerabilities Using Open-source Data
    \titlenote{This work is sponsored by the Intelligence Advanced Research Projects Activity (IARPA) in the Office of the Director of National Intelligence (ODNI) under Air Force Contract FA8702-15-D-0001. The United States Government is authorized to reproduce and distribute reprints for governmental purposes notwithstanding any copyright annotation hereon. The views and conclusions contained herein are those of the authors and should not be interpreted as necessarily representing the official policies or endorsements, either expressed or implied, of IARPA or the United States Government.}}



%
%
%
%

\numberofauthors{4} 
%
\author{
%
%
\alignauthor Benjamin L. Bullough \\
       \affaddr{MIT~Lincoln~Laboratory}\\
\alignauthor Anna K. Yanchenko \\
       \affaddr{MIT~Lincoln~Laboratory}\\
\alignauthor Christopher L. Smith \\
       \affaddr{MIT~Lincoln~Laboratory}\\
\and  
\alignauthor Joseph R. Zipkin \\
       \affaddr{MIT~Lincoln~Laboratory}\\
}



\maketitle
\begin{abstract}
Each year, thousands of software vulnerabilities are discovered and reported to the public.  Unpatched known vulnerabilities are a significant security risk.  It is imperative that software vendors quickly provide patches once vulnerabilities are known and users quickly install those patches as soon as they are available.  However, most vulnerabilities are never actually exploited.  Since writing, testing, and installing software patches can involve considerable resources, it would be desirable to prioritize the remediation of vulnerabilities that are likely to be exploited.  Several published research studies have reported moderate success in applying machine learning techniques to the task of predicting whether a vulnerability will be exploited.  These approaches typically use features derived from vulnerability databases (such as the summary text describing the vulnerability) or social media posts that mention the vulnerability by name.  However, these prior studies share multiple methodological shortcomings that inflate predictive power of these approaches.  We replicate key portions of the prior work, compare their approaches, and show how selection of training and test data critically affect the estimated performance of predictive models.  The results of this study point to important methodological considerations that should be taken into account so that results reflect real-world utility.
\newline
\end{abstract}

%
%
%

%
%

%
%
\printccsdesc



\section{Introduction}
Thousands of software vulnerabilities are discovered and reported to the public every year.  Once a vulnerability becomes public, the odds of it being exploited increase drastically \cite{bilge-before}, making unpatched disclosed vulnerabilities a significant security risk.  Software vendors are generally quick to provide patches once vulnerabilities are known, and vigilant users are quick to install those patches as soon as they are available.  However,  writing, testing, and installing patches can involve considerable resources, requiring organizations to prioritize vulnerabilities based on some notion of risk.  Because most software vulnerabilities are never actually exploited \cite{nayak-vulnerabilities}, it would be desirable to consider the probability that they will be exploited as part of the risk assessment process.

Several previous studies have characterized the exploitation of disclosed vulnerabilities.  Bilge and Dumitras \cite{bilge-before} found that ``after vulnerabilities are disclosed publicly, the volume of attacks exploiting them increases by 5 orders of magnitude''.  They also found that ``exploits for 42\% of all vulnerabilities employed in host-based threats are detected in field data within 30 days after the disclosure date.''  This result illustrates the risk posed by unpatched software vulnerabilities, the need for software vendors and users to quickly provide and install patches and the impact of a failure to patch.

However, Nayak et. al. \cite{nayak-vulnerabilities} find that only 15\% of the known vulnerabilities in a set of popular software products (i.e. MS Windows, MS Office, Internet Explorer, Adobe Reader) are ever exploited ``in the wild'' (i.e. among computers used by the general public \cite{in_the_wild}).  They also find that a small subset of vulnerabilities contribute disproportionately to attacks.  This finding suggests that not all vulnerabilities carry the same risk and thus software makers and system administrators could benefit from a method for prioritizing their response to disclosed software vulnerabilities.

Frei et al. \cite{frei-ecosystem} study the lifecycle of vulnerabilities including creation, discovery, disclosure, exploit availability, patch availability and patch installation.  They describe the major players in a ``security ecosystem'' and how the motives of the different players result in different orderings of the milestones in the vulnerability lifecycle.  Their empirical results show that in most years over 70\% of all exploited vulnerabilities have exploits available at the time of disclosure.  

While there are many sources of data on software vulnerabilities a few deserve special mention due to their ubiquity in research and practice.  The Common Vulnerabilities and Exposures (CVE) dictionary, maintained by the MITRE Corporation, is an authoritative source of identifiers for publicly known information security vulnerabilities \cite{cve-mitre}.  The National Vulnerability Database (NVD) is a ``repository of standards based vulnerability management data'' produced by the U.S. National Institute of Standards and Technology (NIST) \cite{nvd-nist}.  The NVD is a superset of the CVE dictionary and is automatically synchronized with it.  The Common Vulnerability Scoring System (CVSS) is an ``open framework for communicating the characteristics and impacts of IT vulnerabilities'' \cite{first-cvss-guide}.   While CVSS has several components, the CVSS Base Score is the most widely used.  The CVSS Base Score is given on scale from 0 to 10, calculated from a vector of categorical variables that reflect the characteristics of the vulnerability.   

For lack of a better method, the CVSS Base Score is commonly used as an indicator of risk and this practices has been codified in some industry security standards \cite{allodi-casecontrol}.  However, Allodi and Massacci show that the CVSS base score alone is a very poor indicator of whether a vulnerability is likely to be exploited in the wild \cite{allodi-casecontrol}.  They find that the existence of a proof-of-concept exploit and the existence of an exploit traded in cybercrime black markets are much more indicative of the risk of exploitation.

\section{Related Work}
\label{sec:related_work}

Several studies have attempted predict with machine learning techniques whether disclosed software vulnerability will be exploited.  Bozorgi et. al. \cite{bozorgi-exploits} use features  derived from the Open Source Vulnerability Database (OSVDB) and CVE to train a support vector machine (SVM) classifier.  As ground truth they use the OSVDB ``Exploit Classification'' status, coding as positive any entries listed as ``available, rumored or private exploit''.   Their dataset contains vulnerabilities disclosed from 1991--2007.  They report classification accuracy approaching 90\% in offline experiments using a resampled, balanced data set.  They also experiment with online exploit prediction, where classifiers are repeatedly trained with all vulnerabilities seen up to a point in time, and find that classification accuracy stabilizes near 85\%.  Interestingly, 73\% of the vulnerabilities in their data set are labeled as exploited (N=13,765), which contrasts with other surveys and datasets \cite{allodi-casecontrol, nayak-vulnerabilities, edkrantz-said, sabottke-twitter} that show most vulnerabilities are not exploited.

Edkrantz and Said \cite{edkrantz-said} apply several machine learning algorithms to the task of classifying which vulnerabilities will be exploited.  They use vulnerability records from the NVD as a source of features.  They use the presence of an exploit in Exploit Database (EDB) \cite{exploit-db} that references a CVE-ID as a source of ground truth for their study.  (Exploit Database is an archive of software exploits maintained by the company Offensive Security.)  In their overall dataset, which covers the years 2005 through 2014, 27\% of vulnerabilities are labeled as positive (N=55,914).  They report 83\% accuracy on a balanced test set using SVM with a linear kernel (with slightly worse results using Random Forest and Na\"{\i}ve Bayes classifiers). 

Building on the work of Bozorgi \cite{bozorgi-exploits}, Sabottke et al. \cite{sabottke-twitter} use linear SVM classifiers to predict which vulnerabilities will be exploited, but instead of using features generated from the vulnerability description text in the NVD database, they substitute the text of Twitter posts (tweets) that reference a particular CVE-ID.  In addition to predicting public ``proof-of-concept'' exploits that are found in databases such as EDB, they consider prediction of ``real-world'' exploits, based on Symantec anti-virus and intrusion detection signatures, and private proof-of-concept exploits, based on Microsoft's Exploitability Index.  While their data is restricted to the one-year period from January 2014 to January 2015, it is notable that the class ratios are starkly different than in Bozorgi.  While the majority of vulnerabilities in the Bozorgi data are labeled as exploited, only 6.2\% of vulnerabilities in Sabottke's dataset have corresponding private proof-of-concept exploits and only 1.3\% have real-world exploits (N=5865).

\section{Challenges of Realistic Exploit Prediction and System Evaluation}
Machine learning techniques for classification present a compelling approach for the task of predicting which software vulnerabilities will be exploited due to their ability to discover complex patterns within high-dimensional data.  However, care must be taken in selecting training and test data and generating features so that the results will be representative of performance when used for online prediction to aid enterprise network defenders.  We consider several issues that critically affect the evaluation of predictive models and offer suggestions for evaluation that will more readily translate to real-world performance.  These challenges are realistic class balance, appropriate seperation of training and test data, representative and realistic test data and meaningful benchmarks.

In their work surveying the field of intrusion detection, Sommer and Paxson \cite{sommer-ml} examine why machine learning approaches have not been widely deployed despite extensive academic research.  They describe a number of challenges that make this domain particularly difficult, including the difficulty of evaluation, and offer guidelines that they hope will improve the utility of research.  Exploit prediction involves many of the same challenges and adds the additional difficulty of forecasting future events rather than detecting ongoing attacks.  In a similar vein, Rossow et al. describe many of the pitfalls that can limit the correctness and realism of experiments using malware \cite{rossow2012prudent}.  While the context is different from that of exploit prediction, several of the issues they discuss echo findings in this work, including the appropriate seperation of training and evaluation data, removing moot samples, the appropriate balance of classes and use of real-world data.

\begin{table*} [t]
\centering
\renewcommand{\arraystretch}{2}
\begin{tabular}{ | l | c | p{13cm} |}
\hline
Metric & Formula & Description \\ \hline
Accuracy & $ \frac{TP + TN}{TP + TN + FP + FN} $ & Fraction of predictions that are correct \\ \hline
Precision & $ \frac{TP}{TP + FP} $ & Fraction of positive predictions that are correct \\ \hline
Recall & $ \frac{TP}{TP + FN} $  & Fraction of cases with the positive condition that are correctly predicted \\ \hline
F1 & $ 2 * \frac{precision * recall}{precision + recall}  $ & Harmonic mean of Precision and Recall \\ \hline
\end{tabular}
\captionof{table}{Classification performance metrics.  The following abbreviations are used: TP - true positives, TN - true negatives, FP - false positives, FN - false negatives}\label{tab:metrics}
\end{table*}

A variety of metrics are commonly used to evaluate the performance of a classifier, such as accuracy, precision, recall and F1 score.  These are summarized in \autoref{tab:metrics}. Accuracy is the fraction of predictions that are correct.  While it is good to have high accuracy, its drawback as a performance metric is that it fails to distinguish between different types of prediction errors.  Conflating the different types of error can be misleading when their costs are different or when the class ratios are highly imbalanced.  For example, when the classes are highly imbalanced, it becomes increasingly trivial to achieve high prediction accuracy by simply predicting the majority class in every case.

Using precision and recall together distinguishes between the impact of errors that are false positives from those that are false negatives.  Precision is defined as the fraction of positive predictions that are correct and highlights the effect of false positives.  Recall is defined as the fraction of cases with the positive condition that are correctly predicted and highlights the effect of false negatives.  

The F1 metric is the harmonic mean of precision and recall and provides a single number that summarizes the performance of a classifier.  As there is an inherent trade-off between precision and recall, most classifiers can be tuned to select an operating point that is optimized for a particular application.  Precision-recall curves (e.g. \autoref{fig:class_balance_single}) summarize graphically this trade-off for the range of possible values of recall.

\subsection{Class Imbalance}
\label{subsec:class-imbalance}

The ratio of class labels is an important consideration in training and evaluating a classifier.  It is generally more difficult to build a classifier that performs well on both precision and recall when the class ratio is highly imbalanced.  Performance measured on an artificially balanced dataset will generally not be representative of performance when the classes are imbalanced.  Therefore, it is essential to evaluate a classifier on a test set that is representative of the true class ratios; otherwise, the results will be misleading.

\subsection{Train/Test Split}
\label{subsec:train-test-split}

A common practice when developing machine learning models is to split a labeled dataset into two parts, one for training the model and another that is reserved exclusively for evaluating the performance of the model.  A typical method is to partition the data randomly.  This is a best practice when the samples can be assumed to be mutually independent and drawn from the same probability distribution; however, this assumption is not strictly true in the case of exploit prediction.  While some violations of this assumption may be acceptable, it can be problematic in the case of predictive analytics to train models based on information that is available only in the future.   For example, if a new term is coined in the future to describe something about the positive class, it is not reasonable to assume that a model could learn a feature that accounts for its presence based on past training data.  Temporal intermixing of training and test data violates this common sense assumption, and allows information about the future to ``leak'' into the model.  While this could be rationalized as simply ignoring the effects of concept drift, accounting for the impact of concept drift is a critical aspect of evaluating any candidate predictive model.

\subsection{Representative and Realistic Data}
\label{subsec:realistic-data}

In addition to the issues mentioned above, it is important that the data represent the task faithfully in other respects.  In the case of exploit prediction, the goal is to predict whether a vulnerability will be exploited in the future.  However, many vulnerabilities are known to be exploited at the time of disclosure.  Predicting whether such vulnerabilities will be exploited is not a meaningful task and these cases should not be included as test samples.  Similarly, it is important to consider whether a realistic system would be operating in a batch mode, where many historical examples are considered at one time, or an online mode, where each sample is classified as it arrives.  Some features that can be calculated retrospectively in a batch setting use information that might not be available in an online setting, which is more representative of real predictive analytics applications.  For example, if vulnerabilities were to be classified at the time of disclosure, information about the elapsed time between record creation and any subsequent record modifications would not be available.

\subsection{Meaningful Benchmarks}
\label{subsec:meaningful-benchmarks}

Reporting that a particular technique achieves a certain level of performance on a representative task is more useful when given in the context of the next best alternative.  Only by doing so can actual progress be measured. Comparing an approach to a na\"{\i}ve model only shows that the approach is not na\"{\i}ve, not that it advances the state of the art in a meaningful way

\subsection{Examples from Related Work}
\label{subsec:examples}

Much of the prior work in exploit prediction takes an approach at odds with one or more the aforementioned principles.  Bozorgi \cite{bozorgi-exploits} and Edkrantz \cite{edkrantz-said} each evaluate their methods using artificially balanced datasets, inflating performance metrics.  Edkrantz \cite{edkrantz-said} and Sabottke \cite{sabottke-twitter} each randomly split training and test data, allowing the temporal intermixing of ``future'' and ``past'' data in the training and test sets.  Many of the most influential features in the model used by Bozorgi \cite{bozorgi-exploits} are derived from information that would not be available in an online setting (though they do present results from a simulated online experiment).  Sabottke \cite{sabottke-twitter} evaluates the utility of text-based features derived from social media, but never compares those results to using text features derived from the vulnerability summary that is readily available at the time of disclosure.

\section{Method}

In this study, we follow the outlines of prior work to build a model to predict which software vulnerabilities will be exploited.  However, our focus is on illustrating the impact of the previously discussed challenges.  To this end, we systematically vary the approach of selecting training and test examples and generating features to highlight the effects on prediction performance.  

While our approach to training the baseline models follows many of the common aspects of previous work, we are unable to precisely duplicate  any particular study due to the unavailability of data or a lack of detailed knowledge about the methods used.  (For example, OSVDB dataset is no longer publicly accessible.) However, we find that our baseline approach achieves results similar enough to be comparable, and we focus our attention on evaluating the effects of controlled changes to the model building and evaluation methodology.

\begin{table}[!bb]
\begin{center}
\begin{tabular}{|p{1.8cm}|p{2.6cm}|r|r|} \hline
\multicolumn{2}{|c|}{ Name} &  Type & Count \\ \hline\hline
  \multirow{7}{*}{CVSS} & Access Vector & C & 3 \\ \cline{2-4}
  & Access \mbox{Complexity} & C &  3\\ \cline{2-4}
  & Authentication & C & 3 \\ \cline{2-4}
  & Confidentiality Impact & C & 3\\ \cline{2-4}
  & Integrity Impact & C  & 3 \\ \cline{2-4}
  & Availability \mbox{Impact} & C &  3\\ \cline{2-4}
  & Score & N & 1\\ \cline{1-4}

 \multirow{4}{*}{CPE} & Hardware Count & N & 1\\ \cline{2-4}
  & OS Count & N & 1\\ \cline{2-4}
  & Applications Count & N & 1\\\cline{2-4}
  &  List & T & 2000\\ \cline{1-4}
  
   \multirow{7}{*}{CWE} & ID & C & 3\\ \cline{2-4}
   & Name & T& 51\\\cline{2-4}
   & Description & T & 178\\\cline{2-4}
   & \mbox{Number of} \mbox{Parents} & N & 1\\\cline{2-4}
   & Parents & C & 30\\\cline{2-4}
   & Parents' Names & T & 29\\\cline{2-4}
   & Parents' \mbox{Descriptions} & T & 108\\\cline{1-4}
  
   \multirow{4}{*}{References}& Number & N & 1\\\cline{2-4}
   & Type & T & 3\\\cline{2-4}
   & Source & T & 42\\ \cline{2-4}
   & URL & T & 2000\\\cline{1-4}
   Summary & Summary & T & 2000\\\cline{1-4}

\end{tabular}
\captionof{table}{Summary of NVD features, types and counts.  The following abbreviations are used for the feature types:
N - Numeric, C - Categorical, T - Text.
CVSS: Common Vulnerability Scoring System, CPE: Common Platform Enumeration, CWE: Common Weakness Enumeration.}
\label{tab:feat-type}\end{center}
\end{table}

\begin{figure}
\centering
\includegraphics[height=2in, width=3in]{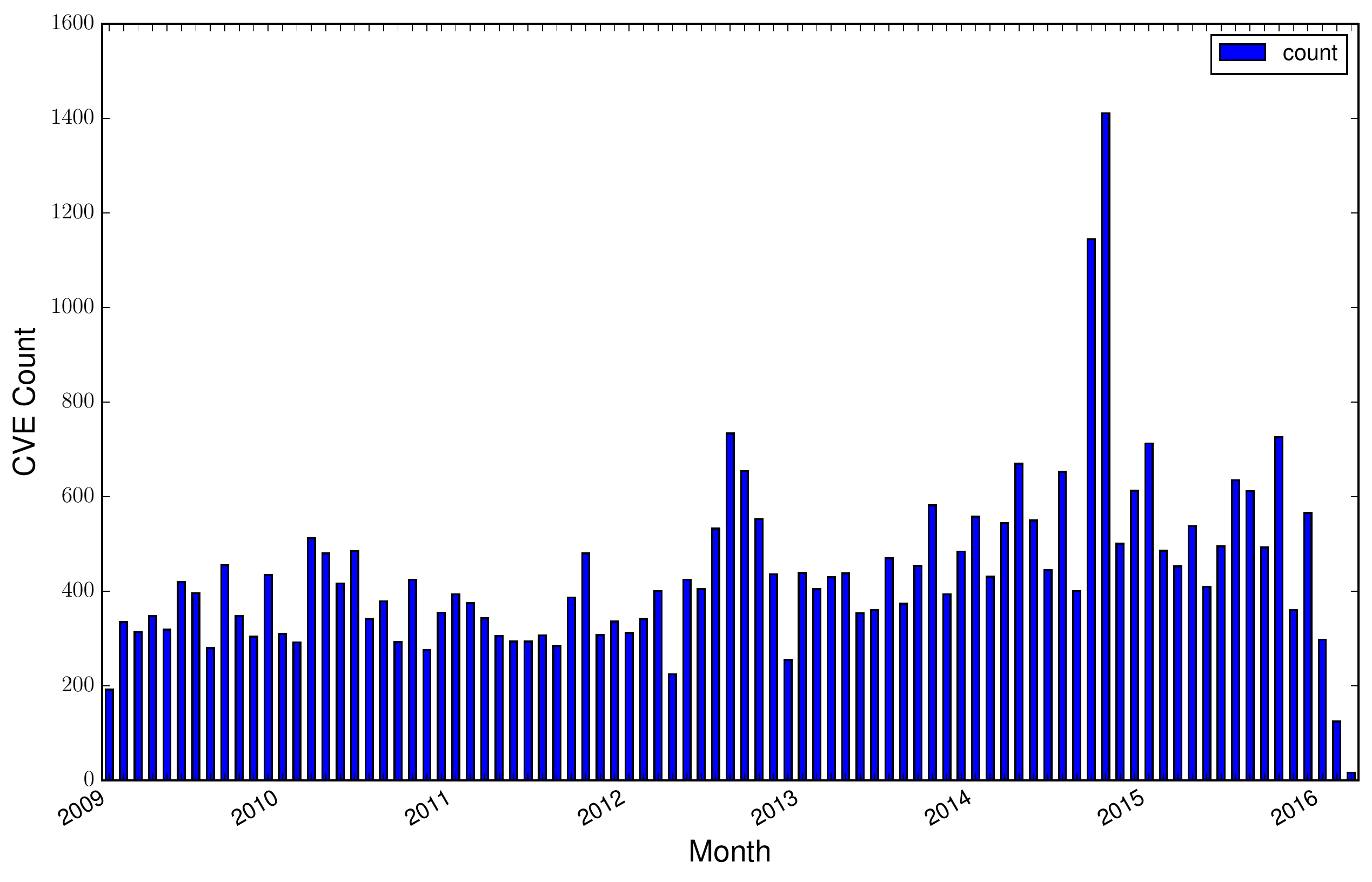}
\caption{Vulnerability disclosures added to the National Vulnerability Database (NVD) by month}
\label{fig:monthly_cve_count_plot}
\end{figure}

We obtained data about published vulnerabilities from the NVD, which indexes vulnerabilities by their CVE-ID  \cite{cve-mitre}.  Our dataset contained 38,129 vulnerabilities published between 2009 and 2015. \autoref{fig:monthly_cve_count_plot} shows the distribution of the disclosure dates over time.  Each record includes a descriptive text summary for the vulnerability, scores and metrics from CVSS, information about affected products and vendors, the category of vulnerability based on the Common Weakness Enumeration (CWE) system, and URLs to other reference sources.  \autoref{tab:feat-type} contains a list of all data features and their type (i.e. numeric, categorical or text). 

The numeric CVSS score vulnerability is calculated from the values assigned to the feature's Access Vector, Access Complexity, Authentication, Confidentiality Impact, Integrity Impact, and Availability Impact.  Access Vector can take on the values ``requires local access'', ``adjacent network accessible'' or ``network accessible''.  Access Complexity can be rated ``high'', ``medium'' or ``low''.  Authentication can take on the values ``requires multiple instances of authentication'', ``requires single instance of authentication'' or ``requires no authentication''.   Confidentiality Impact, Integrity Impact, and Availability Impact can be scored as ``none'', ``partial'' or ``complete''.  These numbers are used to calculate the overall, numeric CVSS score (see \cite{first-cvss-guide}).

Text features that had multiple entries for the same CVE-ID, (i.e. the Vulnerable Systems List, Reference Types, Reference Sources, Reference URLs, CWE Names of Parents, and CWE Descriptions of Parents) were concatenated together into one longer entry for that specific feature. The counts of hardware, operating systems, and applications affected were aggregated from the Vulnerable Systems List. Each CWE has a name and description and there is a hierarchy of CWEs, with CWE-IDs lower on the tree more specific than those above.

We merged the NVD data with data from the Exploit Database by CVE-ID. This data was then further pre-processed, according to feature type. Numerical features were standardized to have zero mean and unit variance, categorical features were represented using binary one-hot encodings, and text features were represented using a bag-of-words model. Each group of text features was individually modeled using a separate term frequency-inverse document frequency (TF-IDF) model, with a maximum of 2000 terms.  (We found there was minimal benefit in using more than 2000 terms.)  Stop words were removed from all text features and the terms ``Exploit-DB'', ``Milw0rm'', and ``OSVDB'' were removed from the Reference Source. Reference URLs that contained any of these three terms were completely removed. These three sources are all databases detailing exploits and they were excluded so as to not leak information about already known exploits into the model.  While the presence of these terms in the domain would be an obvious indicator that the vulnerability had been exploited, other tokens associated with the URL structure for each site could be equally telling, thus we remove the entire URL before extracting features.

For our ground truth labels, we used the presence of a proof-of-concept (POC) exploit in the Exploit Database (EDB) \cite{exploit-db} to choose whether to mark a CVE as exploited or not. CVE references to EDB are collected in a reference map at \cite{exploitdb-cve-refmap} (2319 unique CVEs link to EDB); however, many more EDB entries contain links to CVE-IDs that are missing from this list. To compile a complete list, we used a web scraper to check every EDB entry for a CVE link to generate a full EDB--CVE mapping. This full mapping gave us a list of 19,864 unique CVE entries with POC exploits present in the EDB.  Of these, 6470 were within our target date range.

After splitting our data into training and test sets and pre-processing our features, we trained a support vector machine (SVM) classifier with a linear kernel on the training data.  We used the trained classifier to generate predictions on the test set.  Rather than make a hard classification decision for each test sample and produce single point estimates of precision, recall, or accuracy, we use the raw output of the decision function to plot interpolated precision and recall plots, which show the entire range of precision-recall performance that would be achievable with the classifier depending on the selected threshold.

\begin{center}
\begin{tabular}{|p{4.5cm}|r|r|}\hline 
  Name &  Type & Count \\ \hline\hline
  \# tweets & N & 1 \\ \hline
  \# with high friend count & N & 1 \\ \hline
  \# with high follower count & N & 1\\ \hline
  \# retweets & N & 1 \\\hline
  \# favorited & N & 1\\ \hline
  Ave \# hashtags & N & 1 \\ \hline
  Ave \# URLs & N & 1 \\ \hline
  Ave \# mentions & N & 1 \\ \hline
  \# verified & N & 1 \\ \hline
  Ave account Age & N & 1 \\ \hline
  Body & T & 2000\\ \hline

\end{tabular}
\captionof{table}{Summary of Twitter-derived features, types and counts.  The following abbreviations are used for the feature types:
N - Numeric, T - Text.}
\label{tab:twitter-feat-type}
\end{center}

For our experiments with Twitter data we collected tweets from the Twitter API that contained the term ``CVE'' and selected those tweets containing a valid CVE-ID.  Our Twitter collection covered the first 5 months of 2016.  We then considered only those CVE records for which there was at least one tweet in this time period (2712 in total).  Because our collection system only collects tweets containing the query term ``CVE'', and these can be expected to account for much less than 1\% of all Twitter traffic, we can reasonably assume that our corpus contains all of the matching tweets over the collection time period, despite the 1\% rate limit imposed by the Twitter API \cite{twitter-api}.  Our approach is very similar to Sabottke.  Although our data covers a different time period, we find that the number of unique CVE-IDs in our corpus relative to the collection duration is proportional to that of Sabottke.

\autoref{tab:twitter-feat-type} describes the features, derived from \cite{sabottke-twitter}, that were used in these experiments in addition to the CVSS features from \autoref{tab:feat-type}.  The statistical features are calculated from the collection of tweets that reference a particular CVE-ID.  The lexical features are based on the concatenated text from the bodies of the tweets.  As before we use the presence of the CVE-ID in EDB to assign the class label.

\section{Results}
\label{results}

\subsection{Experiment 1: Class Imbalance}

\begin{figure}
\centering
\includegraphics[height=2in, width=3in]{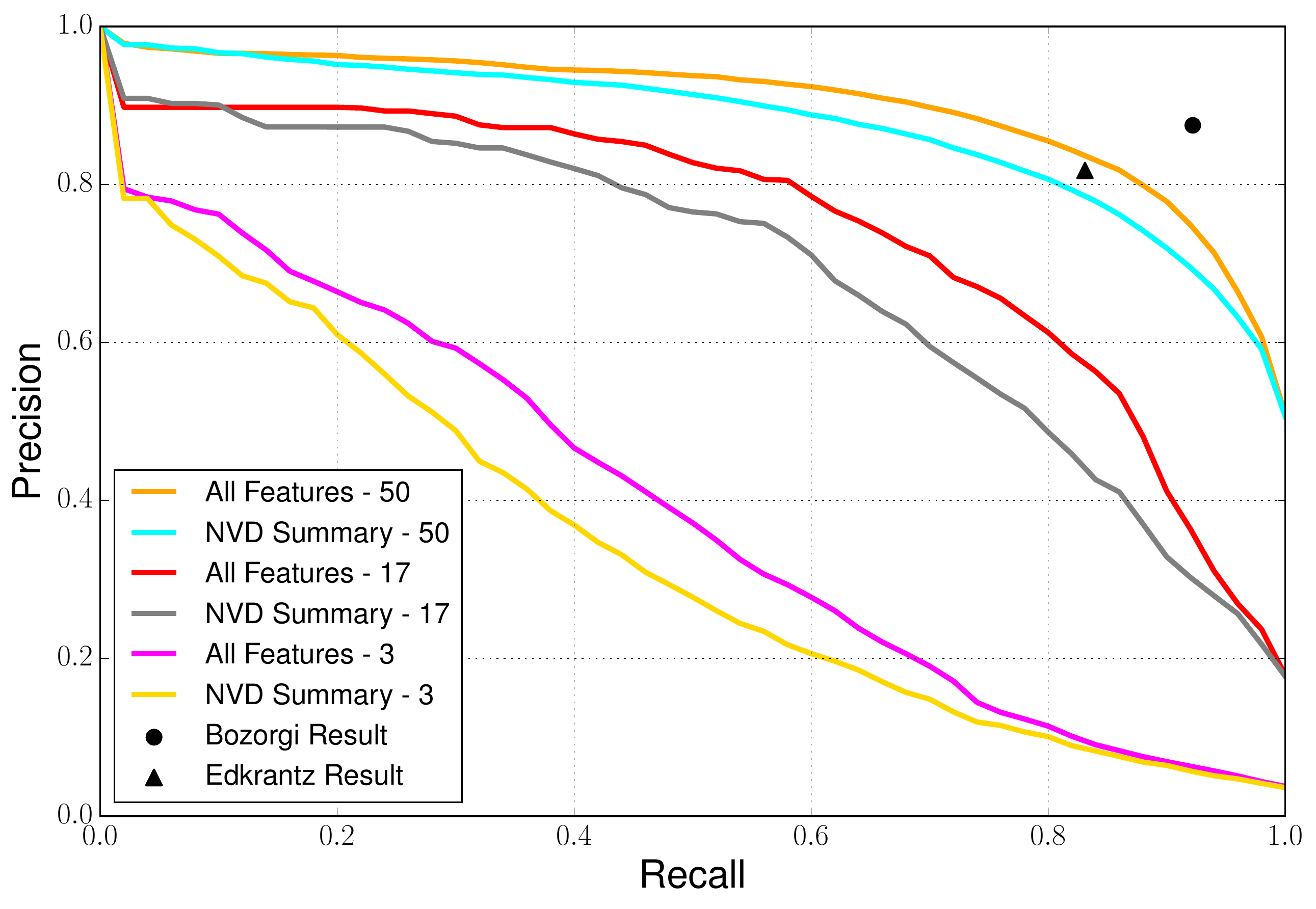}
\caption{Precision-recall plots showing effect of different feature sets and class ratios on classification performance.}
\label{fig:class_balance_single}
\end{figure}

To see the effects of class imbalance, we manipulated the class ratio of our datasets by resampling from our original data set, of which 17\% of the vulnerabilities were exploited (positive class).  To make our results comparable to prior work, we randomly split the data between training and test sets, where 18\% of the data was in the test set.  The results in  \autoref{fig:class_balance_single} and \autoref{tab:class_balance_single} show that performance was significantly boosted by artificially generating balanced datasets.  Likewise, performance dropped further if the imbalance became more severe, which may be realistic, as we discuss in \autoref{sec:related_work}.  When the classes were balanced, performance exceeds the results reported by Edkrantz \cite{edkrantz-said} and approaches the results reported by Bozorgi \cite{bozorgi-exploits}.

\begin{center}
 
\begin{table*} [t]
\centering
\begin{tabular}{|l|r|r|r|r|r|r|}
\hline

& \multicolumn{3}{|c|}{All Features} & \multicolumn{3}{|c|}{NVD Summary} \\
\hline
& 50\% \mbox{Positive} & 17\% \mbox{Positive} & 3\% \mbox{Positive} & 50\% 
\mbox{Positive} & 17\% \mbox{Positive} & 3\% \mbox{Positive} \\
 
\hline
 Accuracy  &                         0.841 &                         0.904 &                        0.968 &                        0.809 &                        0.895 &                       0.967 \\  Precision &                         0.841 &                         0.743 &                        0.566 &                        0.811 &                        0.736 &                       0.582 \\  Recall    &                         0.845 &                         0.651 &                        0.362 &                        0.802 &                        0.577 &                       0.171 \\  F1        &                         0.843 &                         0.694 &                        0.442 &                        0.807 &                        0.647 &                       0.264 \\\hline
\end{tabular}
\captionof{table}{Performance metrics for All Features and only NVD Summary  for various class balances.}\label{tab:class_balance_single}
\end{table*}
\end{center}

\subsection{Experiment 2: Randomized vs. Temporal Train/Test Split}

\begin{figure}
\centering
\includegraphics[height=2in, width=3in]{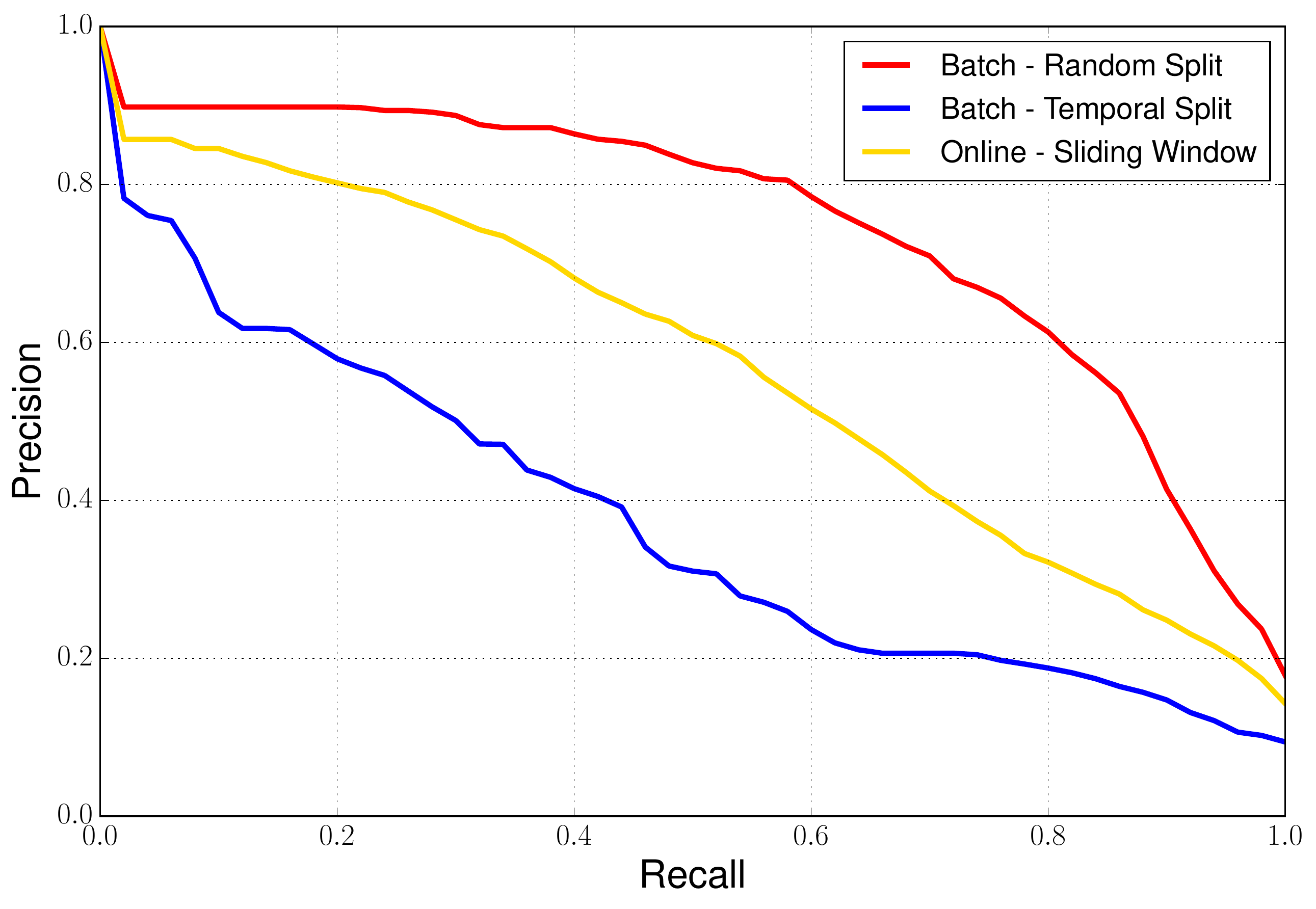}
\caption{Precision-recall plots comparing the method of splitting training and test sets.}
\label{fig:train_test_split}
\end{figure}

\begin{center}
 
\begin{tabular}{|p{1.3cm}|p{1.7cm}|p{1.7cm}|p{1.7cm}|}
\hline
             &   Batch \mbox{Random} Split &   Batch \mbox{Temporal} Split &   Online \mbox{Sliding} Window \\
\hline

 Accuracy  &                  0.904 &                    0.902 &                     0.874 \\  Precision &                  0.742 &                    0.466 &                     0.553 \\  Recall    &                  0.651 &                    0.342 &                     0.564 \\  F1        &                  0.693 &                    0.394 &                     0.558 \\\hline
\end{tabular}\captionof{table}{Performance metrics for various training/test splits.}\label{tab:train_test_split}
\end{center}

\begin{figure*}
\centering
\includegraphics[height=2.4in, width=5in]{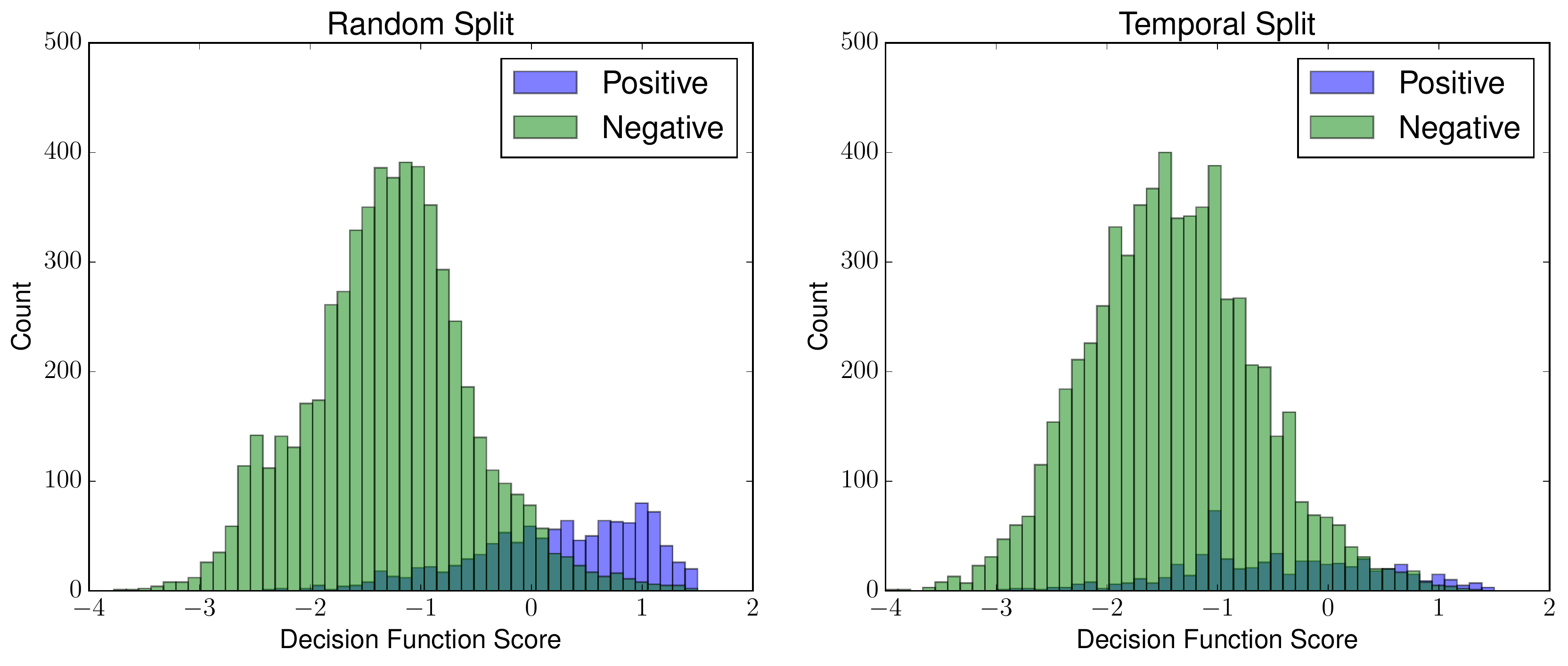}
\caption{Class separation for one realization of a random split (left) and a temporal split (right) for the NVD Summary text only and using the top 2000 terms.}
\label{fig:class_sep-train_test_split}
\end{figure*}

To see the effect of temporal intermixing of the training and test data, we compared splitting the data by date (where all vulnerabilities disclosed after January 1, 2015, were placed in the test set), to randomly selecting an equivalent portion of the data as the training set.  Since all vulnerabilities published after January 1, 2015, accounted for about 18\% of the data, we also used 18\% of the data in the test set for the random split approach.  We additionally evaluated an online approach where we trained on the first year of data and evaluated on the subsequent six months of data.  We then added the test data to the training set, retrained the model, and evaluated it on the following six months, and so on. \autoref{fig:train_test_split} and \autoref{tab:train_test_split} demonstrate that random splitting of the training and test data substantially inflate the performance of the predictive model.  The results using a sliding-window online approach are between the batch approaches, suggesting it is a reasonable approach that minimizes the negative effects of concept drift while properly restricting training data to events that have occurred in the past.

We note that the fraction of positive cases in the temporally split test set is only 9.3\%, which is less than the 16.7\% of positive cases in the randomly split test set.  This shows that the class ratio in our data is not stable over time, and it means that some of the decrease in performance is attributable to the class imbalance being somewhat more extreme when we use only the more recent data for evaluation.  However, given the quantitative results from our experiment with class imbalance, this amount of change in the class ratio is not enough to account for the majority of the effect on performance.  To further examine the difference in performance between random and temporal splits, we looked at the class separation achieved by the classifier decision function learned with each type of split. \autoref{fig:class_sep-train_test_split} shows significantly more class separation in the random split case, as compared to the temporal split case.  The classifier was not able to distinguish between the positive and negative classes in the temporal spilt case as successfully as it was for the random split, leading to the difference in performance.

\subsection{Experiment 3: No Pre-exploited Vulnerabilities}

\begin{figure}
\centering
\includegraphics[height=2in, width=3in]{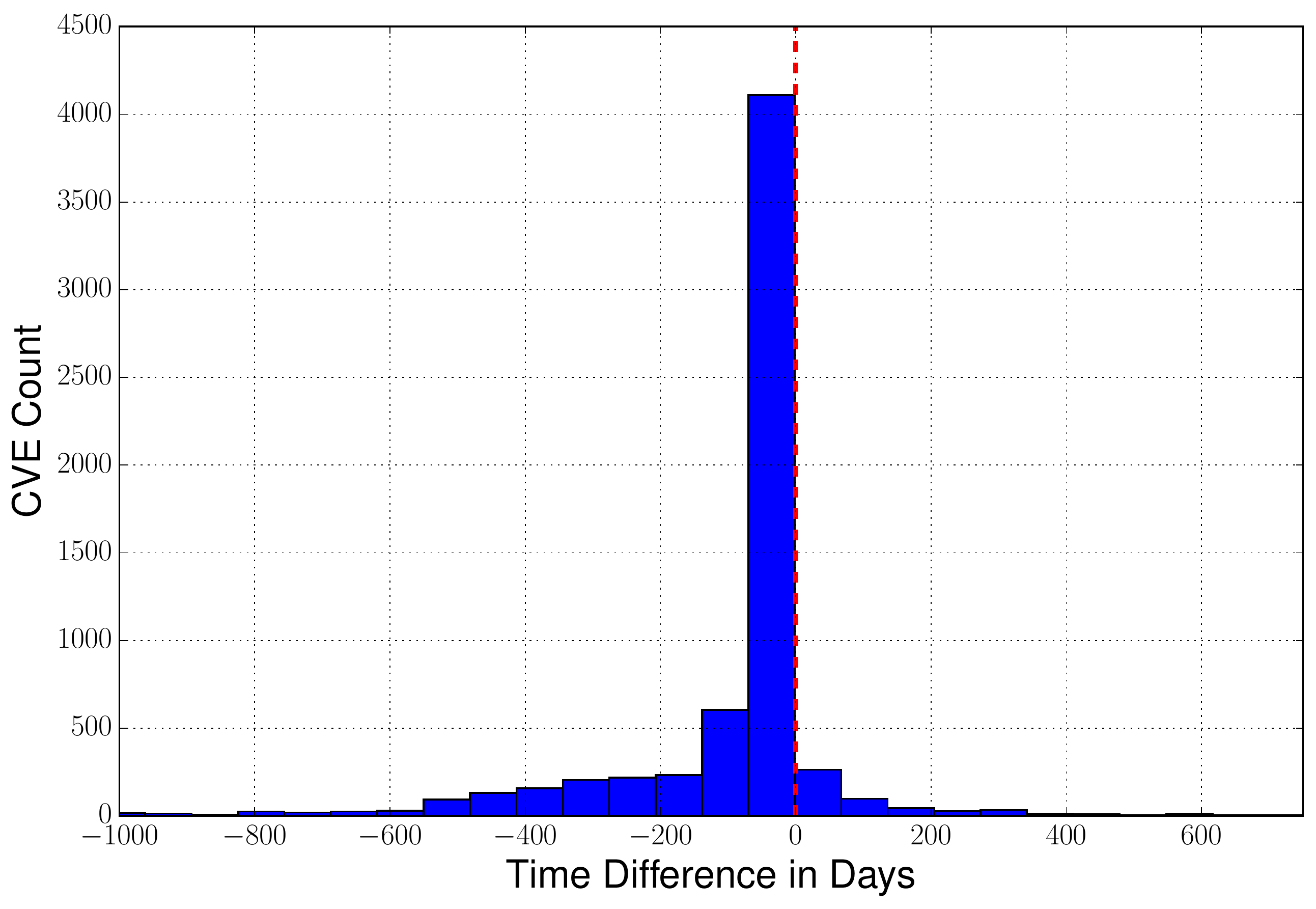}
\caption{Distribution of the difference in days between vulnerability disclosure and exploit publication}
\label{fig:exploit_time_difference_hist}
\end{figure}

\begin{figure}
\centering
\includegraphics[height=2in, width=3in]{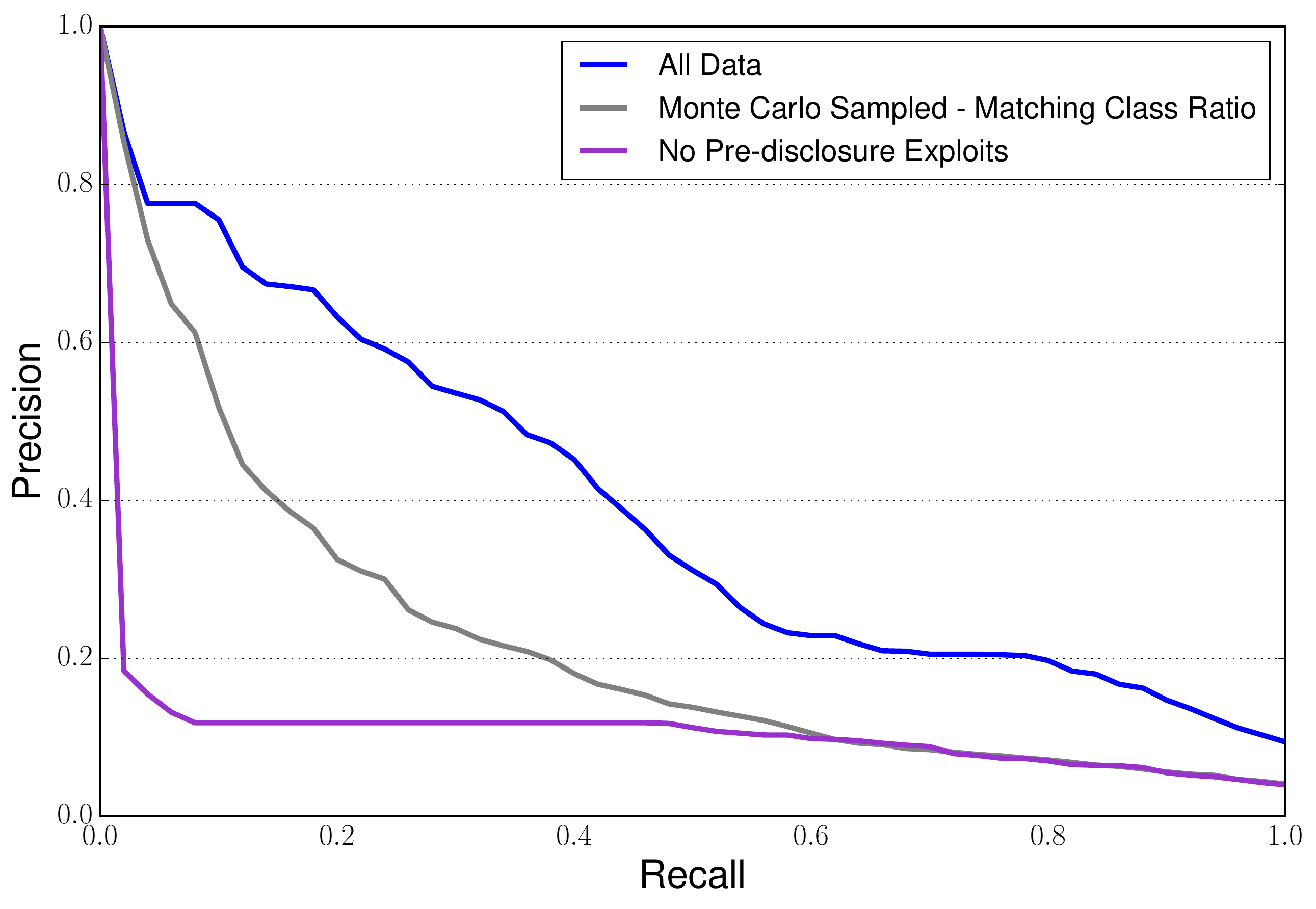}
\caption{Precision-recall plots illustrating the effect of removing from the dataset any vulnerabilities with exploits published prior to disclosure.  The effect of resampling to achieve a similar class ratio is shown for comparison.}
\label{fig:no_exp}
\end{figure}

It is important to measure the performance of a predictive model on a task that is representative of the realistic situation.  In many cases, exploits are published before or concurrently with a vulnerability disclosure.  \autoref{fig:exploit_time_difference_hist} shows a histogram of the difference in days between vulnerability disclosure in the NVD database and exploits that target those vulnerabilities being published in EDB.   Most vulnerabilities in the NVD that are exploited have already been exploited by the time they are disclosed.  Since a simple database search can determine that these have been exploited, they are not really candidates for prediction.   When we excluded these examples from the dataset we found that performance was drastically reduced to a level not much better than guessing.  Since this also exacerbates the class imbalance, we compared this performance to a dataset where a random sample of the positive cases was removed so that the class ratio was equivalent to the former case.  We see in \autoref{fig:no_exp} and \autoref{tab:no_exp} that while the effect of increasing the class imbalance is substantial, performance is still much better than when all of the previously exploited cases are removed.

Since the task of predicting the exploitation of disclosed vulnerabilities is premised on the idea that exploits are commonly developed based on the vulnerability disclosures (``zero-day'' exploits being the obvious exception), it was surprising to find such a large fraction of cases where the vulnerability publication date is later than the earliest publication date of a proof-of-concept exploit that references that vulnerability.  \autoref{sec:example_vulns} describes how it is possible for an exploit to reference a vulnerability by its CVE-ID prior to its publication and provides examples of specific CVEs.  While an exploit does exists at the time of vulnerability disclosure in many cases, a limitation of our data is that it does not indicate when any database updates that link the two may have occurred, which would better indicate public knowledge about the exploited status of a vulnerability.  If such updates do not always occur very soon after the vulnerability publication, then our approach of removing all vulnerabilities from the dataset where the exploit is published prior to disclosure might seem overly aggressive.  On the other hand, if we assume that any vulnerabilities can be quickly matched to existing exploits, then using a simple database lookup on past exploits as a classifier would achieve 100\% precision at 59.3\% recall on the full test set while ignoring any future exploit development.  This suggests that any added value must come by improving performance on the case where exploits are not yet developed.

\begin{center}
 
\begin{tabular}{|p{1.5cm}|p{1.3cm}|p{1.7cm}|p{2.4cm}|}
\hline
             &   \mbox{All Data} Points &   \mbox{Matching} Class Ratio &  \mbox{No Pre-disclosure} Exploits \\
\hline

 Accuracy  & 0.909 &      0.959 &                        0.956 \\  Precision & 0.519 &      0.458 &                        0.171 \\  Recall    & 0.334 &      0.105 &                        0.027 \\  F1        & 0.406 &      0.170 &                        0.046 \\\hline
\end{tabular}\captionof{table}{Performance metrics for the full dataset, excluding vulnerabilities 
which were exploited before being published in NVD, and resampled to achieve a similar 
class ratio.}\label{tab:no_exp}
\end{center}

\subsection{Experiment 4: Twitter Text vs. NVD Summary Text}

\begin{figure}
\centering
\includegraphics[height=2in, width=3in]{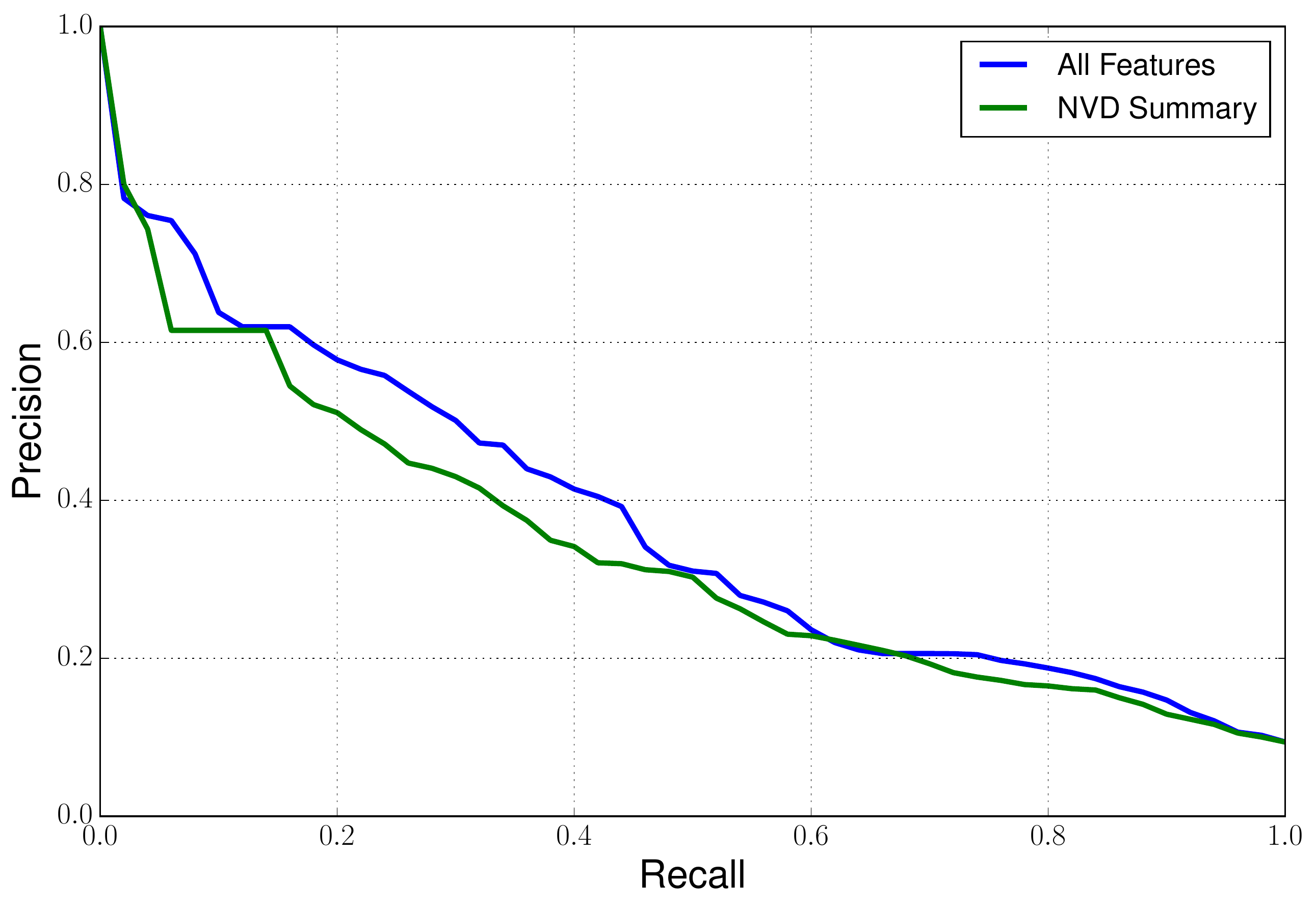}
\caption{Precision-recall plots comparing the classification performance using all features with performance using only lexical features derived from the NVD summary text.}
\label{fig:feature_selection}
\end{figure}

\begin{center}
 
\begin{tabular}{|l|r|r|}
\hline
            &   All Features &   NVD Summary \\
\hline

 Accuracy  &          0.902 &         0.897 \\  Precision &          0.466 &         0.426 \\  Recall    &          0.342 &         0.311 \\  F1        &          0.394 &         0.359 \\\hline
\end{tabular}\captionof{table}{Performance metrics for All Features and NVD Summary features only.}
\label{tab:feature_selection}
\end{center}                                        

\begin{figure}
\centering
\includegraphics[height=2in, width=3in]{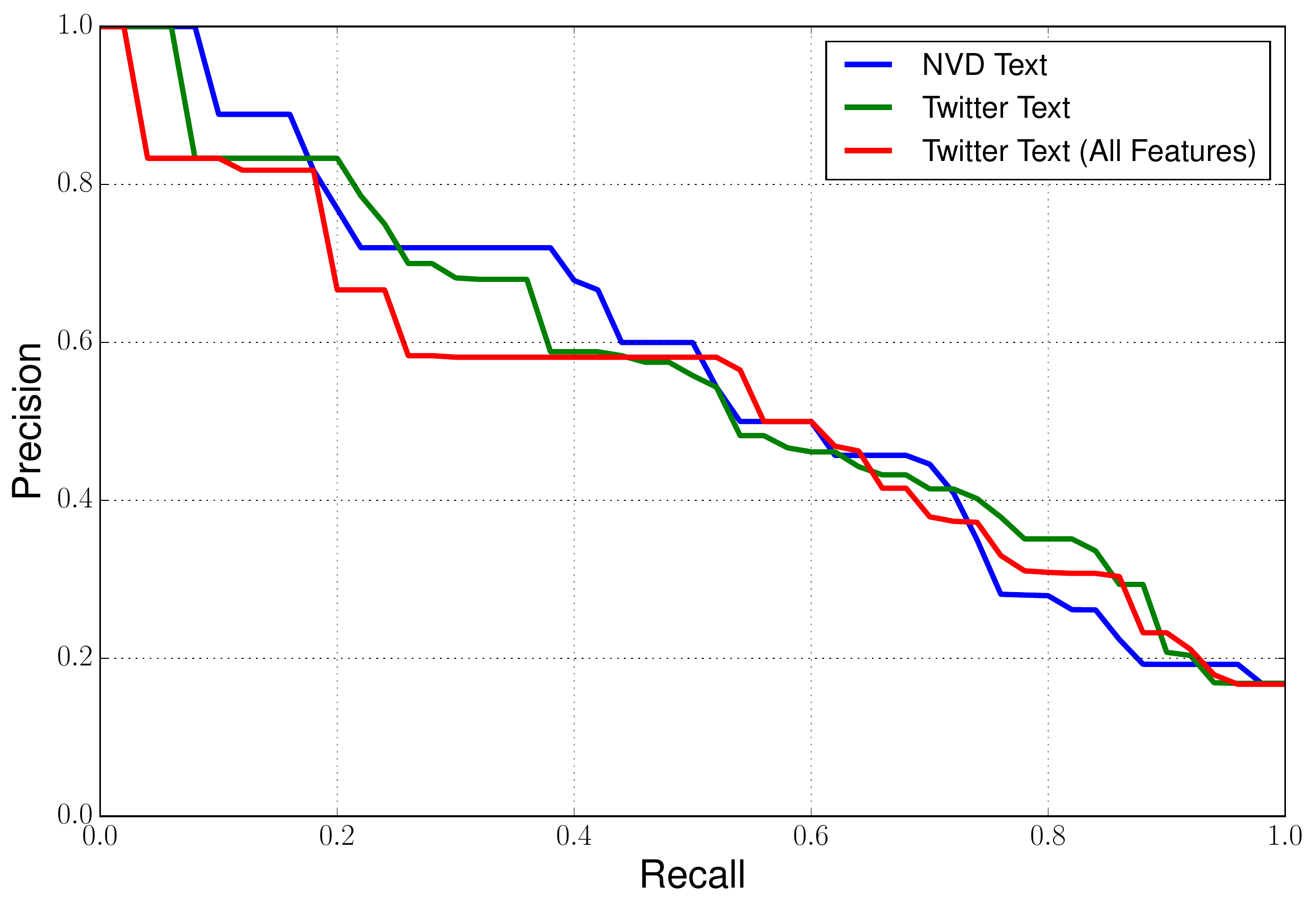}
\caption{Precision-recall plots comparing classification performance using features derived from Twitter to features derived from the NVD summary text.}
\label{fig:twitter_vs_nvd}
\end{figure}

The lexical features derived from the NVD summary text account for almost all of our classifier's performance.   \autoref{fig:feature_selection} and \autoref{tab:feature_selection} show that all the other features combined provide a very modest improvement, mainly when precision is emphasized over recall.

Sabottke investigates using features derived from Twitter to predict whether software vulnerabilities will be exploited.  They combine lexical features from the text of tweets mentioning CVEs with other statistical features from the Twitter data and features from the NVD database.  While using social media appears to produce promising results, they do not compare their results to a baseline approach of using the lexical features in the NVD summary text.   \autoref{fig:twitter_vs_nvd} compares our results using features derived from the NVD database (summary and CVSS scores) to our results using features derived from Twitter along with the CVSS scores.  (The results using NVD text in this experiment differ from \autoref{fig:feature_selection} due to using a different range of time.)  We show the results using only the lexical features as well as augmented with most of the statistical features used by Sabottke.  In both cases, the model using Twitter features does not show an improvement over the baseline model.  

Collecting information about vulnerabilities from Twitter involves a significant investment in data processing capabilities, delays the time to make a prediction until the Twitter community generates sufficient tweets, and can only be applied to vulnerabilities that generate social media interest.  The open nature of the Twitter platform also poses a greater risk of an adversary attempting to poison the defenders training data, though this was investigated by Sabottke \cite{sabottke-twitter} who found the impact would be modest under several models of adversarial interference.  Given these considerations the marginal benefits of using social media would need to be sufficient to justify the costs.  Our results suggest that such a benefit does not exist.

\section{Discussion}

The results of the four experiments described in the previous section show that subtle changes in the evaluation process can have a profound impact on the estimated performance of a predictive system.  Making more realistic choices about class balance, how to split training and test data and what constitues a realistic example for prediction lowered the F1 score of our classifier from 83\% to 4\%.  While the appropriate choices for each of these dimensions is highly dependent on the prediction task and training data, this example illustrates why it is important to make realistic decisions for each of them in order to get performance estimates that are likely to translate to a pratical setting.

Our assessment of the value of predictions based on NVD data is consistent with the results of Zhang et al. \cite{zhang-nvd}, who attempt to use NVD data to predict the time until the next \textit{undiscovered} vulnerability is reported in a particular piece of software.  They conclude that data in the NVD generally has poor predictive capability.  One reason that NVD data may have poor predictive power is the presence of substantial concept drift (as seen in \autoref{fig:train_test_split}), which makes it difficult to generalize about future vulnerabilities based on descriptions of past vulnerabilities.

We also find limited use for social media in predicting whether software vulnerabilities will be exploited.  While we focus on the issue of performance relative to an established baseline, Tufekci \cite{tufekci-socialmedia} describes several methodological pitfalls, such as over-reliance on a single platform, unrepresentative samples, and field effects that should also be considered when using social media big data for predictive analytics. 

While models based on the open source database information examined in this study led to poor prediction performance, more accurate predictions might be possible by exploiting better sources of data.  Developing richer or higher-fidelity databases about vulnerabilities and exploits, perhaps by collaboration with organizations that hold large amounts of private data, is one approach that exploits the potential of big data for security analytics.  Another avenue might be to use static or dynamic analysis of vulnerable code or binaries to obtain greater insight into the fundamental exploitability of a particular vulnerability or its utility as a target for exploitation.  Exploring such approaches is left for future work. 

A limitation on effectiveness of any predictive system operating in an adversarial environment is that its predictions may influence future events, thereby undermining the apparent accuracy of its predictions.  For example, if predictions about software exploitation become widely known and relied upon, attackers may intentionally shift their focus to another less defended target.  Such an offensive strategy presumably imposes some cost on attackers by pushing them towards less desirable avenues of attack, so there is still some net benefit to the defender for deploying the predictive system.  This limitation exists regardless of the mechanism used for prediction, but systems based on machine learning may be quicker to adapt to evolving attacker strategy than systems based on manually developed rules.  Exploring the co-evolution of attacker and defender strategies and the impact on the effectiveness of defensive systems is also left to future work.

\section{Conclusions}

In this study we identify several methodological considerations that need to be accounted for during training and evaluation of predictive models that will be used in practical settings.  Previous studies that have investigated predicting exploits of disclosed software vulnerabilities have failed to account for some or all of these issues.  We show empirically the impact of violating these principles on the task of exploit prediction.  We conclude that models of software vulnerability exploitation based on NVD database entries and social media alone are unlikely to have enough predictive power to be useful in practice.

\section{Acknowledgments}
We thank Jeremy Blackthorne and Clark Wood for their helpful feedback on drafts of this paper. 

%
\bibliographystyle{abbrv}
\bibliography{iwspa01-bullough}  
%
%

\appendix
\section{Vulnerabilities with Pre-existing Exploits}
\label{sec:example_vulns}

The median difference in time between exploit publication and vulnerability publication in our dataset is 8 days, but there are many individual cases where this gap is much larger.  One might ask how it is possible to reference a vulnerability by its CVE-ID prior to its publication.  One explanation is that CVE-IDs can be allocated prior to publication in the NVD database.  This would allow a security researcher who finds a vulnerability to reserve a CVE-ID, publish their proof of concept exploit and then publish the vulnerability, in that order.  (The researcher may also disclose the vulnerability privately to the vendor prior to public disclosure in accordance with responsible disclosure practices as described by Frei. \cite{frei-ecosystem})  An example that seems to fit this pattern is CVE-2015-4414, a directory-traversal vulnerability in a WordPress audio plugin that allows an attacker to read arbitrary files on the system.  The CVE-ID for this vulnerability was reserved on 7 June 2015.  An exploit (EDB-ID: 37274) that references this CVE-ID was published on EDB on 12 June 2015.  The vulnerability was published on the NVD on 17 June 2015.

The record of an exploit might also be updated after its initial publication to reference a vulnerability that is subsequently published.  This may be because the exploit author did not disclose the vulnerability and thus didn't know the CVE-ID prior to the vulnerability publication.  It may be that a newly discovered vulnerability is readily exploited using an existing exploit with only a trivial modification.  An example of the later case is CVE-2011-5289, published in the NVD on 31 December 2014, which exploits the SaveDecrypted method in the aTube Catcher ActiveX control to write arbitrary files.  This vulnerability is referenced in EDB by an exploit (EDB-ID: 6963), which was published on 3 November 2008.  This exploit also references an earlier vulnerability (CVE-2008-5002), published on 10 November 2008.  This vulnerability affects the same ActiveX control and allows an attacker to overwrite arbitrary files via the WriteFile method.

\end{document}